\documentclass[twocolumn]{aa}
\usepackage{graphicx}
\usepackage{natbib}
\usepackage{amsmath}
\bibpunct{(}{)}{;}{a}{}{,}
\def\kms{$\mbox{km s}^{-1}$}
\begin{document}
\title{Line-of-sight velocity distribution corrections for \\ Lick/IDS
  indices of early-type galaxies}


   \author{Harald Kuntschner}

   \offprints{Harald Kuntschner}

   \institute{Space Telescope European Coordinating
  Facility, European Southern Observatory, Karl-Schwarzschild-Str. 2,
  85748 Garching, Germany\\
              \email{hkuntsch@eso.org}
             }

   \date{Received ..., 2004; accepted ..., 2004}

   \abstract{We investigate line-of-sight velocity distribution (LOSVD)
     corrections for absorption line-strength indices of early-type
     galaxies in the Lick/IDS system. This system is often used to
     estimate basic stellar population parameters such as luminosity
     weighted ages and metallicities. Using single stellar population
     model spectral energy distributions by Vazdekis (1999) we find
     that the LOSVD corrections are largely insensitive to changes in
     the stellar populations for old galaxies (age $>3$\,Gyr). Only the
     Lick/IDS Balmer series indices show an appreciable effect, which
     is on the order of the correction itself.  Furthermore, we
     investigate the sensitivity of the LOSVD corrections to
     non-Gaussian LOSVDs. In this case the LOSVD can be described by a
     Gauss-Hermite series and it is shown that typical values of $h_3$
     and $h_4$ observed in early-type galaxies can lead to significant
     modifications of the LOSVD corrections and thus to changes in the
     derived luminosity weighted ages and metallicities. A new, simple
     parameterisation for the LOSVD corrections, taking into account
     the $h_3$ and $h_4$ terms, is proposed and calibrations given for
     a subset of the Lick/IDS indices and two additional indices
     applicable to old ($>$3\,Gyr) stellar populations.
     \keywords{Line: profiles -- Methods: data analysis -- Galaxies:
       abundances -- Galaxies: elliptical and lenticular, cD --
       Galaxies: kinematics and dynamics}
}

   \maketitle
%

\section{Introduction}
Absorption line strength measurements have long been used to
investigate the stellar populations of (early-type) galaxies which are
typically too far away to be resolved into individual stars. The
interpretation of the measurements are performed via stellar population
models which have become more and more sophisticated in their
predictions over the years. For example, recent models
\citep[e.g.,][]{tra00a,TMB03} are able to predict line-strength indices
as a function of age, metallicity {\em and}\/ abundance ratios.  The
success of these studies does not only depend on high signal-to-noise
data and up-to-date model predictions but also on accurate calibrations
of the observed data to a standard system. This is particularly
important when comparing data between different objects such as giant
ellipticals and globular clusters.

There are typically three stages in the calibration process: 1)
choosing a line-strengths system which allows the measurement of the
absorption features of interest in a large range of objects; 2)
correcting the line-strength measurements for instrumental effects and
3) correcting the observed data for differences caused by the objects
themselves, such as the stellar kinematics or emission line
contamination. In this paper we will concentrate on point three and
specifically on the corrections due to different line-of-sight velocity
distributions (LOSVD) in galaxies.

In Section~\ref{sec:line_strengths} we summarize the method of
measuring line-strength indices in the Lick/IDS system while we focus
on the effects of non-Gaussian LOSVDs in Section~\ref{sec:nonGaussian}.
Section~\ref{sec:template} presents the stellar population models we
use and the effects of ages and metallicities in our simulations. In
Section~\ref{sec:new_model} we present the new parameterisation of the
LOSVD corrections applicable to old stellar population and LOSVDs which
can be described by a fourth order Gauss-Hermite series. A literature
comparison for our LOSVD corrections is given in
Section~\ref{sec:lit_comp}. Section~\ref{sec:discussion} gives a brief
discussion of the results while the conclusions are listed in
Section~\ref{sec:conclusion}.

\section{Measuring line-strength indices for early-type galaxies}
\label{sec:line_strengths}
For a purely stellar object the observed spectrum at a given sky
position is the luminosity weighted sum of all stars redshifted
according to their line-of-sight velocities and convolved with the
instrumental resolution. However, one typically assumes that the
observed data can be described by the luminosity weighted mean of a
{\em single}\/ age, {\em single}\/ metallicity and {\em single}\/
abundance ratio model-spectrum convolved with the LOSVD and
instrumental resolution.  Even though this is a gross simplification
one can learn a great deal about the average stellar populations in
galaxies.

\begin{figure}
\resizebox{\hsize}{!}{\includegraphics{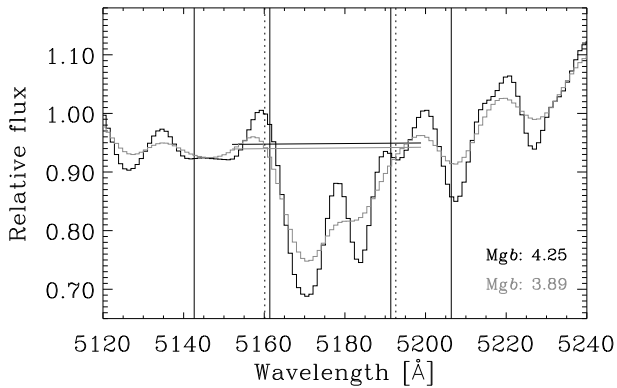}}
   \caption{The effects of velocity broadening are shown for the
     Mg\,$b$\/ region. The black spectrum represents a stellar
     population model of solar metallicity and an age of 12.6\,Gyr at
     an ``instrumental'' resolution of $\sim$4.3\AA~(FWHM). The grey
     line shows a simulated galaxy spectrum with a Gaussian velocity
     dispersion profile of 250\,\kms\/ observed at the same
     instrumental resolution. Overplotted are the bandpasses of the
     Lick/IDS Mg\,$b$\/ index and the resulting pseudo-continua. The
     solid vertical lines show the bandpasses of the two continuum
     bands and the dotted vertical lines indicate the central bandpass
     of the index. The line-strength measured for each spectra is given
     in the figure; the units are \AA.}
           \label{fig:broad}%
\end{figure}

One of the most successful line-strength systems for the interpretation
of early-type galaxies was created by the Lick group
\citep[][]{fab85,wor94b,tra98}.  In the Lick/IDS system,
absorption-line strengths are measured by indices, where a central
feature bandpass is flanked to the blue and red by pseudo-continuum
bandpasses (see Figure~\ref{fig:broad}). The mean height in each of the
two pseudo-continuum regions is determined,
and a straight line is drawn through the midpoint of each one. The
difference in flux between this line and the observed spectrum within
the feature bandpass determines the index. For most absorption features
the indices are expressed in \AA ngstr\"{o}ms of equivalent width. For
broad molecular bands the index is expressed in magnitudes.

A significant LOSVD of stars in galaxies broadens the spectral
features, in general reducing the observed line-strength in the
Lick/IDS system compared to the intrinsic value of the average stellar
population (see Figure~\ref{fig:broad}). The net change in absorption
strength is the combination of several competing effects and depends
on the detailed absorption line composition in the vicinity of the main
index feature. In general there are two ways to affect the observed
index due to the LOSVD of stars: (1) the absorbed flux in the central
index bandpass itself changes and (2) the pseudo-continuum level
changes. Since the continuum of early-type galaxies in the optical
wavelength region is completely eaten away by absorption lines, both
effects play an important role.

For example, the measured absorption strength becomes weaker if stars
in a given galaxy have large enough relative velocities so that part of
the main index absorption feature extends beyond the central index
bandpass.  If the pseudo-continuum bandpasses are close to the index
bandpass then the same effect will additionally lower the
pseudo-continuum. However, if there are significant absorption features
in the pseudo-continuum bandpasses the pseudo-continuum level is lifted
up due to the broadening and therefore the observed index value becomes
larger.

In order to compare the line-strength measurements at different
positions across a galaxy or between different objects, a given index
must be corrected to the value it would have if measured at some
standard LOSVD. Generally, it is convenient to correct all indices to
the values they would have if measured in an object without internal
velocity structure and at a given instrumental resolution which can be
described by a simple Gaussian. In the Lick/IDS system this is a
wavelength dependent value which ranges from approximately 8.4\,\AA\/
to 11.5\,\AA\/ \citep[for details see][]{worott97}.

\begin{figure}
\resizebox{\hsize}{!}{\includegraphics{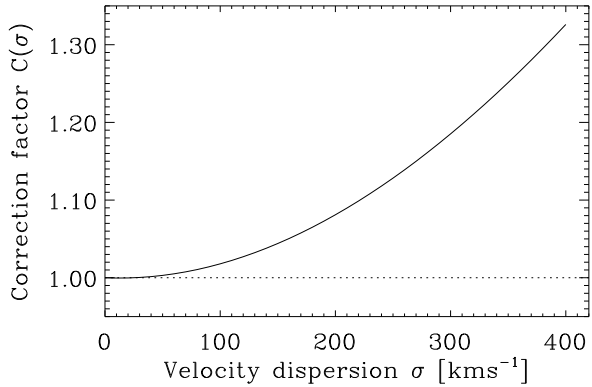}}
  \caption{Simple velocity dispersion correction curve for the
    Mg\,$b$\/ index in the Lick/IDS system and purely Gaussian LOSVDs
    (solid line).}
  \label{fig:velcorr}%
\end{figure}

When correcting Lick/IDS indices for galaxies, typically only the first
moment of the LOSVD, $\sigma$, is taken into account, hence the
corrections have been referred to as velocity dispersion corrections
\citep[e.g.,][]{dav93}. An example of such a correction
factor\footnote{$\mathrm{Index}_{\mathrm{corrected}} =
  \mathrm{Index}_{\mathrm{observed}} \times \mathrm{C}(\sigma)$} as a
function of $\sigma$ is given in Figure~\ref{fig:velcorr}. One can see
that the correction strength is a rising function of the velocity
dispersion $\sigma$ and reaches 12.6\% at $\sigma = 250$\,\kms\/ for
the Mg\,$b$\/ index.  Other indices show similar curves, however with a
varying magnitude of the correction factor \cite[e.g.,][]{kun00}.

\section{Effects of non-Gaussian velocity profiles}
\label{sec:nonGaussian}
\begin{figure}
\resizebox{\hsize}{!}{\includegraphics{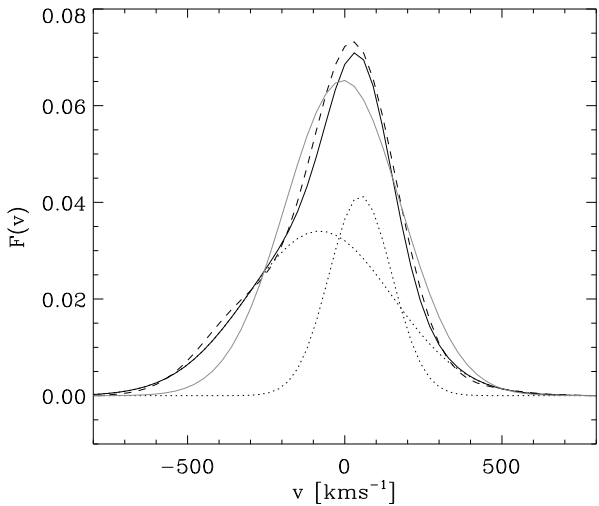}}
  \caption{Example LOSVD (solid black line) constructed by a double
    Gaussian. The two individual Gaussian components are shown as
    dotted lines. The best fitting fourth order Gauss-Hermite
    parametrisation is plotted with a dashed line, while the best
    fitting purely Gaussian LOSVD is given as grey solid line.}
  \label{fig:losvd}%
\end{figure}

With the availability of high quality measurements of significant
non-Gaussian terms of the LOSVD \citep[e.g.,][]{ben94} it is necessary
to investigate the possible effects on the line-strength measurements.
We start with a LOSVD constructed by a double Gaussian (see
Figure~\ref{fig:losvd}). This LOSVD was designed to show a significant
non-Gaussian shape, which qualitatively is representative for a number
of early-type galaxies \citep[see e.g.,][]{sco95}. A double Gaussian of
the following form was used

\begin{equation}
  F(v) = \sum_{j=1}^{2} I_j \exp \Bigg( \frac{(v-V_j)^2}{2 \sigma_j^2} \Bigg)
\end{equation} 
with parameters $I_1 = 0.028$, $I_2 = 0.053$, $V_1 = 50.0$\,\kms,
$V_2=-80.0$\,\kms, $\sigma_1=100.0$\,\kms, and $\sigma_2=230.0$\,\kms.
This set of parameters was constrained to give a LOSVD with a mean
velocity close to zero.

In order to test the influence on the line-strength indices, we use a
single stellar population spectral energy distribution (hereafter SED)
of 10\,Gyr and solar metallicity \citep{vaz99} with known
line-strength.  We convolve this spectrum with our test LOSVD to
simulate an early-type galaxy. The LOSVD corrections for our test LOSVD
are determined by comparing the line-strengths measured on the
simulated galaxy with those from the un-convolved spectrum. These
corrections need to be compared with the classical procedure of
measuring a simple velocity dispersion (ignoring the non-Gaussian
components) from our test spectrum and calculating a correction factor
with the help of calibrations as shown in Figure~\ref{fig:velcorr}.
The best fitting Gaussian LOSVD is represented by $V=-5.8$\,\kms\/ and
$\sigma = 183.4$\,\kms\/ (see solid, grey line in
Figure~\ref{fig:losvd}).

For the Mg\,$b$\/ index we infer a LOSVD correction factor of 1.089 for
the true test LOSVD while for the best fitting Gaussian we infer a
factor of 1.068 -- a difference of 2.1\% to the true correction factor.
It is clear from this example that line-strength indices in the
Lick/IDS system can be sensitive to non-Gaussian LOSVDs and that this
warrants a more systematic investigation into the LOSVD corrections.

In practise it is more convenient to describe the non-Gaussian terms of
the LOSVD by the Gauss-Hermite parametrisation introduced by
\citet{mf93} and \citet{ger93}. The best fit to our example LOSVD using
a fourth order Gauss-Hermite series is obtained with parameters
$V=0.2$\,\kms, $\sigma=173.8$\,\kms, $h_3=-0.119$, and $h_4=0.082$ (see
dashed line in Figure~\ref{fig:losvd}). This parameterisation gives
generally very good fits to observed LOSVDs and can therefore be used
conveniently to parametrise otherwise complex LOSVDs. In our example
the correction factor would be underestimated by only 0.2\% when using
the Gauss-Hermite parametrisation.

Using the integral-field-unit {\tt SAURON}, \citet{em04} report
measurements of $h_3$ and $h_4$ terms in a representative sample of
early-type galaxies. Most of the targets show 2-dimensional structure
for the $h_3$ and $h_4$ terms with values as large as $\pm0.2$.
Although the $h_3$ terms tend to be close to zero in the central
region of early-type galaxies, a non-zero $h_4$ term is frequently
observed. In our example LOSVD both $h_3$ and $h_4$ were non-zero and
it is interesting to see if the line-strength indices show a different
sensitivity to them.
 
\begin{figure*}
\centering
\includegraphics[width=17.0cm]{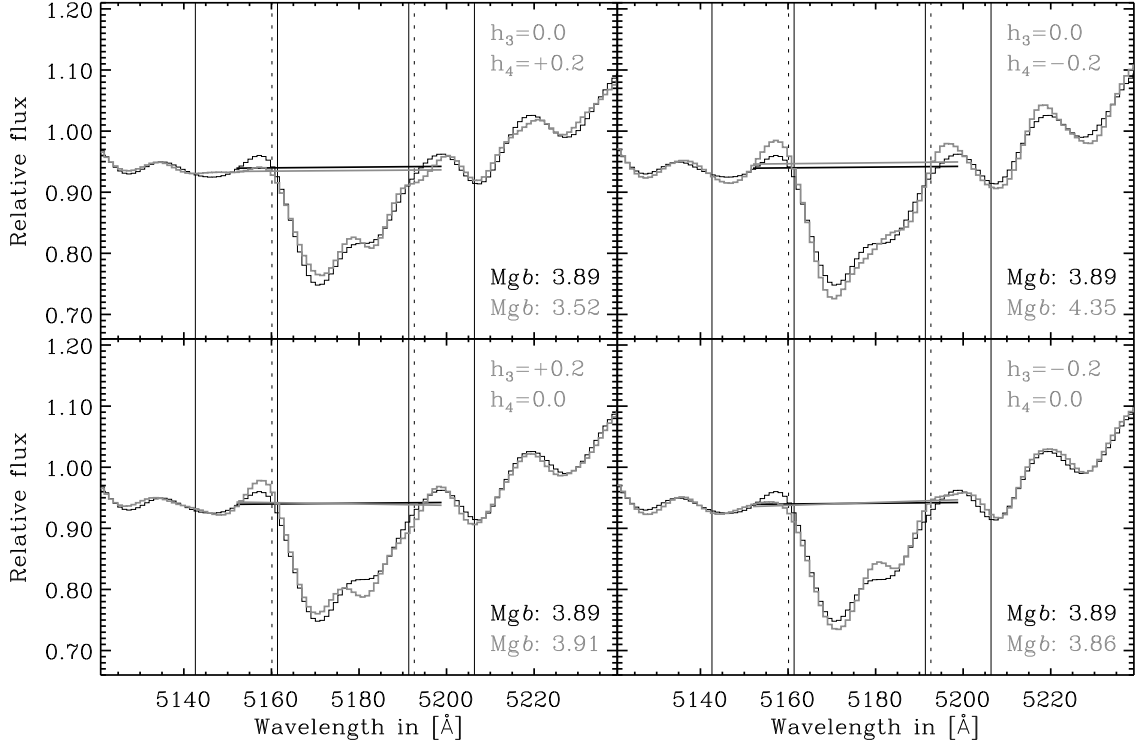}
   \caption{The effects of velocity broadening for non Gaussian
     LOSVDs are shown for the Mg\,$b$\/ region. The black spectrum in
     all panels represents a stellar population model of solar
     metallicity and 12.78\,Gyr at a velocity dispersion of 250\,\kms,
     $h_3 = 0.0$ and $h_4 = 0.0$. The grey spectra simulate a galaxy
     with different LOSVDs characterised by $\sigma = 250$\,\kms, $h_3
     = \pm0.2$ and $h_4 = \pm0.2$ as indicated in the upper right
     corner of each panel. The index measurements are given in the
     lower right corner of each panel; the units are \AA. Overplotted
     are the bandpasses of the Lick/IDS Mg\,$b$\/ index and the
     resulting pseudo-continua.  The solid vertical lines show the
     bandpasses of the two continuum bands and the dotted vertical
     lines indicate the central bandpass of the index.}
   \label{fig:broadh4}
\end{figure*}

In Figure~\ref{fig:broadh4} we show spectra with non-Gaussian LOSVDs
for four different combinations of $h_3$ and $h_4$ and a constant
$\sigma = 250$\,\kms\/ in the Mg\,$b$\/ region. For the Mg\,$b$\/
line-strength index a $h_4 = \pm0.2$ causes changes in the observed
index strength of approximately $\mp10$\% compared to a purely Gaussian
LOSVD of $\sigma = 250$\,\kms. A non-zero $h_3$ term ($\pm0.2$), has
negligible effects ($<1$\%).

\section{Simulations and the effects of different stellar populations}
\label{sec:template}
The main aim of this paper is to provide a simple procedure to correct
for the LOSVD broadening of line-strength indices in the Lick/IDS
system. Since the corrections are sensitive to the detailed absorption
line composition around the features of interest, it is crucial to
investigate these corrections on simulated galaxy spectra which are a
good representation of real observations. It is well known that
early-type galaxies harbour stellar populations which can span a wide
range in ages and metallicities \citep[e.g.,][]{kun98,tra00a}. For this
purpose we have chosen to use a sub-set of 27 single stellar population
(SSP) model spectra from the work of \citet[][]{vaz99}. The models
cover a large range in metallicity ([Fe/H] = -0.68 to +0.2) and age
(1.0 to 17.78\,Gyr). For all models we use an ``unimodal'' IMF type
with Salpeter slope value. The spectra cover two windows of 3820 to
4500\,\AA\/ and 4780 to 5460\,\AA\/ at a resolution of 1.8\,\AA\/
(FWHM), including a large set of Lick/IDS indices (see
Table~\ref{tab:coeff}).

In order to match the Lick system the spectra were first broadened to
the wavelength dependent Lick/IDS resolution \cite[see][]{worott97}.
All spectra were further broadened to velocity dispersions of $\sigma =
40$ to 400~\kms\/ in steps of 40~\kms\/, $h_3=-0.2$ to $+0.2$ in steps
of $0.04$ and $h_4=-0.2$ to $+0.2$ in steps of $0.04$ in order to
simulate the observed LOSVDs of galaxies. In total there are 32\,697
simulated galaxy spectra per spectral window. The set of Lick/IDS
indices \citep{tra98} and two additional indices are then measured for
each simulated galaxy spectrum and correction factors, $C_{j,k}(\sigma,
h_3, h_4)$, are determined such that

\begin{equation}
\begin{split}
C_{j,k}(\sigma, h_3, h_4) &= \\
  I_{j,k}(\sigma=0, h_3=0, h_4=0) & / I_{j,k}(\sigma, h_3, h_4)
\end{split}
\end{equation}
where $I_{j,k}$ is the index measured for the stellar population model
$k$ and index $j$.

For molecular indices (CN$_1$, CN$_2$) and indices which can have
values close to zero (H$\delta_{\mathrm{A,F}}$,
H$\gamma_{\mathrm{A,F}}$) additive corrections, $C_{j,k}(\sigma, h_3,
h_4)$, are determined such that

\begin{equation}
\begin{split}
C_{j,k}(\sigma, h_3, h_4) &= \\
  I_{j,k}(\sigma=0, h_3=0, h_4=0) & - I_{j,k}(\sigma, h_3, h_4)
\end{split}
\end{equation}
where $I_{j,k}$ is the index measured for the stellar population model
$k$ and index $j$. Table~\ref{tab:coeff} lists all the indices
considered in this paper and also indicates whether additive or
multiplicative corrections are used.

\begin{figure}
\resizebox{\hsize}{!}{\includegraphics{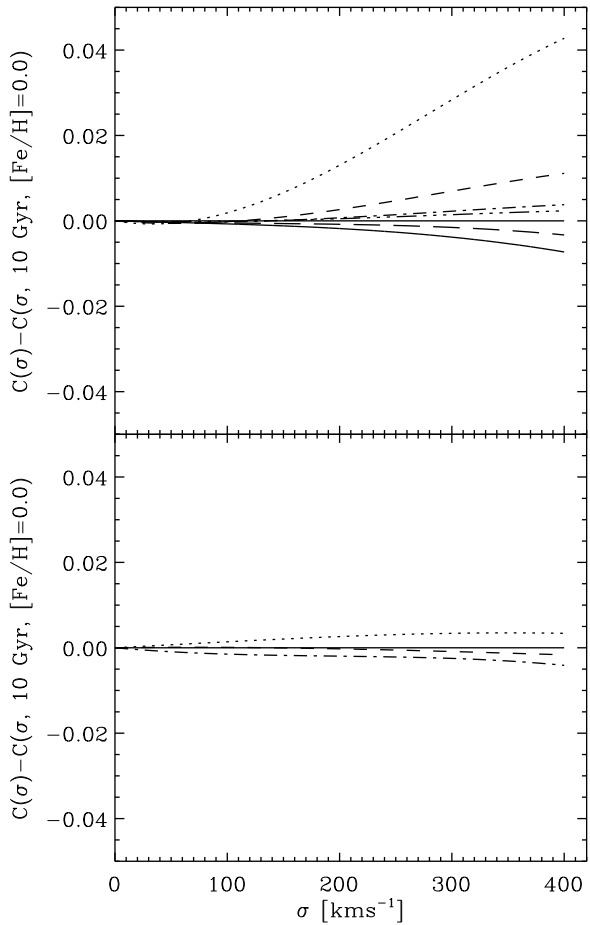}}
  \caption{LOSVD corrections for the Mg\,$b$\/ index as
    function of age and metallicity and purely Gaussian LOSVDs. The
    curve for a stellar population of 10\,Gyr and solar metallicity
    (solid line) has been subtracted from all other curves in order to
    show the details of the deviations. The top panel shows, from the
    top, ages of 1.00, 2.00, 3.16, 5.62, 10.00, 12.59, and 17.78 at
    solar metallicity. The bottom panel shows, from the top,
    metallicities [Fe/H] = 0.20, 0.00, -0.38, -0.68 at an age of
    10\,Gyr.}
  \label{fig:velcorr2}%
\end{figure}

The above measurements on the simulated galaxy spectra yield
corrections for each line-strength index and stellar population model
at a given LOSVD characterised by $\sigma$, $h_3$ and $h_4$. One of the
first questions to ask here is whether the LOSVD correction at a given
LOSVD is a function of input stellar population model, i.e. the age and
metallicity of the simulated galaxy spectrum. Overall, the changes with
age and metallicity are surprisingly small ($<$1\%) as demonstrated in
Figure~\ref{fig:velcorr2} for the Mg\,$b$\/ index.  However, for
stellar populations younger than about 3\,Gyr one can see significant
deviations.

The other indices with multiplicative corrections show a similar
behaviour with average scatters smaller than 1\%. Only the G4300 and
Fe4383 indices show a larger scatter with 1.2\% and 1.8\%,
respectively. Typically, the scatter due to different stellar
population models increases with increasing velocity dispersion. We
note, that the H$\beta_{\mathrm G}$ index, a variant of the Lick/IDS
H$\beta$ index shows a reduced sensitivity to stellar population
changes (error 0.3\% and 0.7\%, respectively).

For indices with additive corrections, the CN$_{1,2}$ indices show very
little scatter with stellar population model, while the higher order
Balmer series indices do show significant scatter. Here the uncertainty
is on the order of the corrections themselves. At first sight this is
an unfavourable situation, however, the corrections are small and the
uncertainties are only a fraction ($<8\%$) of the index change between
a stellar population of 5.62 and 10.0\,Gyr at solar metallicity.
Therefore, the indices are still useful for stellar population studies.
All former uncertainties due to different stellar populations were
derived by using purely Gaussian LOSVDs and are quoted as average
1$\sigma$ uncertainties over the full range of velocity dispersions.

Here in this paper the aim is to provide simple to use LOSVD
corrections for {\it old}\/ stellar populations, so we decided to
average the measurements for all ages greater than 3\,Gyr and all
metallicities covered by the model spectra.

We note, that more accurate LOSVD corrections can be achieved by
fitting an optimal template to the galaxy spectrum in question
\citep[e.g.,][]{cap04}. This fit, obtained in real pixel space, can
provide the necessary parametrisation of an individual LOSVD as well as
the average stellar spectrum. Using this optimal template one can then
determine more accurate LOSVD corrections, especially for galaxies with
young stellar populations.

\section{The new model for LOSVD corrections}
\label{sec:new_model}
Taking into account all simulations with an age of greater than 3\,Gyr
we find that the following models provide a good description of the
corrections. For indices with multiplicative corrections:

\begin{equation}
\label{equ:model_mult}
\begin{split}
C_j(\sigma, h_3, h_4) = 1.0 + \sum_{i=1}^{3} a_{i,j} \sigma^i \\
  + \sum_{i=1}^{2} b_{i,j} \sigma^i h_3 + \sum_{i=1}^{2} c_{i,j} \sigma^i h_4
\end{split}
\end{equation}

For the additive corrections the model is changed to:

\begin{equation}
\label{equ:model_add}
\begin{split}
C_j(\sigma, h_3, h_4) = 0.0 + \sum_{i=1}^{3} a_{i,j} \sigma^i \\
  + \sum_{i=1}^{2} b_{i,j} \sigma^i h_3 + \sum_{i=1}^{2} c_{i,j} \sigma^i h_4
\end{split}
\end{equation}
where $a_{i,j}$, $b_{i,j}$ and $c_{i,j}$ are the correction
coefficients for index $j$ (see Table~\ref{tab:coeff}).

A LOSVD corrected index is then for multiplicative corrections

\begin{equation}
I_{j}^{corr} =  I_{j}^{raw} \times C_{j}(\sigma, h_3, h_4)
\end{equation}

and for additive corrections

\begin{equation}
I_{j}^{corr} = I_{j}^{raw} + C_{j}(\sigma, h_3, h_4)
\end{equation}
where $I_{j}^{raw}$ is the raw index measurement for index $j$ and
$C_{j}(\sigma, h_3, h_4)$ is the LOSVD correction factor.

Table~\ref{tab:coeff} lists the correction coefficients for all indices
investigated in this paper. Table~\ref{tab:coeff_err} gives the average
1$\sigma$ error and the 99.73\% percentile level error for
$C_j(\sigma,h_3,h_4)$. The former errors are derived by averaging over
all considered stellar population models and LOSVDs.

\begin{table*}
   \centering
   \caption[]{LOSVD correction coefficients$^a$ for stellar populations
     with ages $>3$\,Gyr}
   \label{tab:coeff}
   \begin{tabular}{clcrrrrrrr}
     \hline \hline
     $j$ & Name & Type$^{\mathrm{b}}$ & $a_1$ & $a_2$ &$a_3$ & $b_1$ & $b_2$ & $c_1$ &$c_2$\\
     \hline
     1 & H$\delta_{\mathrm A}$  & a &  1.6352E-05& -3.4479E-06&  3.3455E-09&  1.1306E-03& -3.1261E-06& -2.2092E-03&  1.0247E-06\\ 
     2 & H$\delta_{\mathrm F}$  & a & -2.7223E-05&  1.5629E-06& -3.9026E-10&  1.4632E-03& -2.8438E-06&  8.9675E-04&  2.8072E-06\\ 
     3 & CN$_1$                 & a &  4.5496E-07&  5.6976E-08& -6.9482E-11&  3.3855E-05& -1.2322E-07&  3.2924E-05& -3.2799E-08\\ 
     4 & CN$_2$                 & a & -9.1107E-07&  2.0562E-07& -2.2045E-10& -9.1159E-05&  1.4759E-07&  1.3265E-04& -1.2433E-07\\ 
     5 & Ca4227                 & m &  4.1692E-05&  2.6781E-06&  5.7468E-09& -2.6918E-03&  6.5988E-06&  8.9098E-04&  1.4347E-05\\  
     6 & G4300                  & m &  2.3222E-05&  6.3570E-07&  1.6306E-10& -5.6194E-04&  1.9549E-06& -7.4895E-05&  3.7196E-06\\ 
     7 & H$\gamma_{\mathrm A}$  & a &  1.1688E-05&  1.0194E-06& -5.9652E-09& -5.8547E-04& -5.0147E-06&  2.2800E-03& -2.0744E-05\\ 
     8 & H$\gamma_{\mathrm F}$  & a & -1.7727E-05& -4.9682E-07&  2.0945E-09& -1.7022E-03& -4.8060E-07& -1.3818E-04&  3.4325E-06\\ 
     9 & Fe4383                 & m &  1.1509E-05&  2.3090E-06& -3.1740E-10& -9.4865E-04&  5.1610E-08&  9.1133E-04&  4.4302E-06\\ 
     10& H$\beta$               & m &  4.7881E-05& -2.0608E-07&  1.7285E-09&  7.0527E-04&  5.2770E-07& -7.5619E-04&  6.5777E-06\\
     11& H$\beta_G$ $^c$        & m & -3.2132E-05&  1.0408E-06&  4.0073E-10&  7.4397E-04& -1.2992E-06&  4.6158E-04&  3.2367E-06\\
     12& Fe5015                 & m &  8.8838E-06&  3.0684E-06& -2.5073E-09& -1.0867E-03&  2.5696E-06&  1.5059E-03&  6.8840E-07\\
     13& Mg\,$b$                & m & -6.1814E-05&  2.4787E-06& -7.1595E-10& -6.2044E-05& -2.7694E-08&  1.1524E-03&  3.0182E-06\\
     14& Fe5270                 & m &  6.9948E-06&  2.7662E-06& -1.7504E-09&  1.2910E-04&  3.0655E-07&  7.3827E-04&  4.1853E-06\\
     15& Fe527S $^c$            & m & -2.5596E-05&  9.1585E-07&  2.3466E-09&  4.5289E-04& -1.0934E-06& -8.9776E-05&  8.6598E-06\\
     16& Fe5335                 & m & -9.5088E-05&  5.7773E-06& -6.0153E-10&  9.0887E-04& -5.8071E-06&  3.5863E-03&  1.1284E-06\\
     17& Fe5406                 & m & -1.1774E-04&  5.3334E-06&  1.5920E-09&  8.3065E-05&  2.5328E-06&  3.7854E-03&  2.1638E-06\\
        \hline 
      \end{tabular}
  \begin{list}{}{}
  \item[$^{\mathrm{a}}$] The coefficients listed in this table are to
    be used with equations~\ref{equ:model_mult} \& \ref{equ:model_add}
    in order to derive the LOSVD corrections.
     \item[$^{\mathrm{b}}$] This column indicates whether a
       multiplicative (m) or an additive (a) correction is used; see
       equation~\ref{equ:model_mult} \& \ref{equ:model_add}
       respectively.
     \item[$^{\mathrm{c}}$] These indices are not defined in the
       Lick/IDS system. For a definition of H$\beta_G$ see
       \citet{jor97}. For Fe527S, a variant of the Fe5270 index used in
       the {\tt SAURON} survey (Kuntschner et al., in preparation), the
       blue, central and red bandpasses are defined as
       5233.0--5250.000\,\AA, 5256.5--5278.5\,\AA, and
       5285.5--5308.0\,\AA, respectively.
  \end{list}
\end{table*}

\begin{table}
   \centering
   \caption[]{The average errors for LOSVD correction coefficients}
   \label{tab:coeff_err}
   \begin{tabular}{clccc}
     \hline \hline
     $j$ & Name &  Type$^{\mathrm{a}}$ & Error$^{\mathrm{b}}$&99.73\% error$^{\mathrm{c}}$\\
     \hline
     1 & H$\delta_{\mathrm A}$  & a &   0.080&   0.277\\ 
     2 & H$\delta_{\mathrm F}$  & a &   0.027&   0.103\\ 
     3 & CN$_1$                 & a &   0.001&   0.004\\ 
     4 & CN$_2$                 & a &   0.001&   0.004\\ 
     5 & Ca4227                 & m &   0.015&   0.053\\  
     6 & G4300                  & m &   0.013&   0.040\\ 
     7 & H$\gamma_{\mathrm A}$  & a &   0.039&   0.146\\ 
     8 & H$\gamma_{\mathrm F}$  & a &   0.061&   0.207\\ 
     9 & Fe4383                 & m &   0.021&   0.076\\ 
     10& H$\beta$               & m &   0.013&   0.045\\
     11& H$\beta_G$             & m &   0.006&   0.019\\
     12& Fe5015                 & m &   0.006&   0.019\\
     13& Mg\,$b$                & m &   0.006&   0.015\\
     14& Fe5270                 & m &   0.005&   0.015\\
     15& Fe527S                 & m &   0.004&   0.013\\
     16& Fe5335                 & m &   0.012&   0.031\\
     17& Fe5406                 & m &   0.017&   0.044\\
     \hline 
  \end{tabular}
  \begin{list}{}{}
     \item[$^{\mathrm{a}}$] This column indicates whether a multiplicative (m) or an additive (a) correction is used; see equation~\ref{equ:model_mult} and \ref{equ:model_add} respectively.
     \item[$^{\mathrm{b}}$] Average 1$\sigma$ error on the correction factor $C_j(\sigma,h_3,h_4)$.
     \item[$^{\mathrm{c}}$] 99.73\% percentile level error on the correction factor $C_j(\sigma,h_3,h_4)$.
  \end{list}
\end{table}

\section{Literature comparison}
\label{sec:lit_comp}
\begin{figure}
   \centering
   \resizebox{\hsize}{!}{\includegraphics{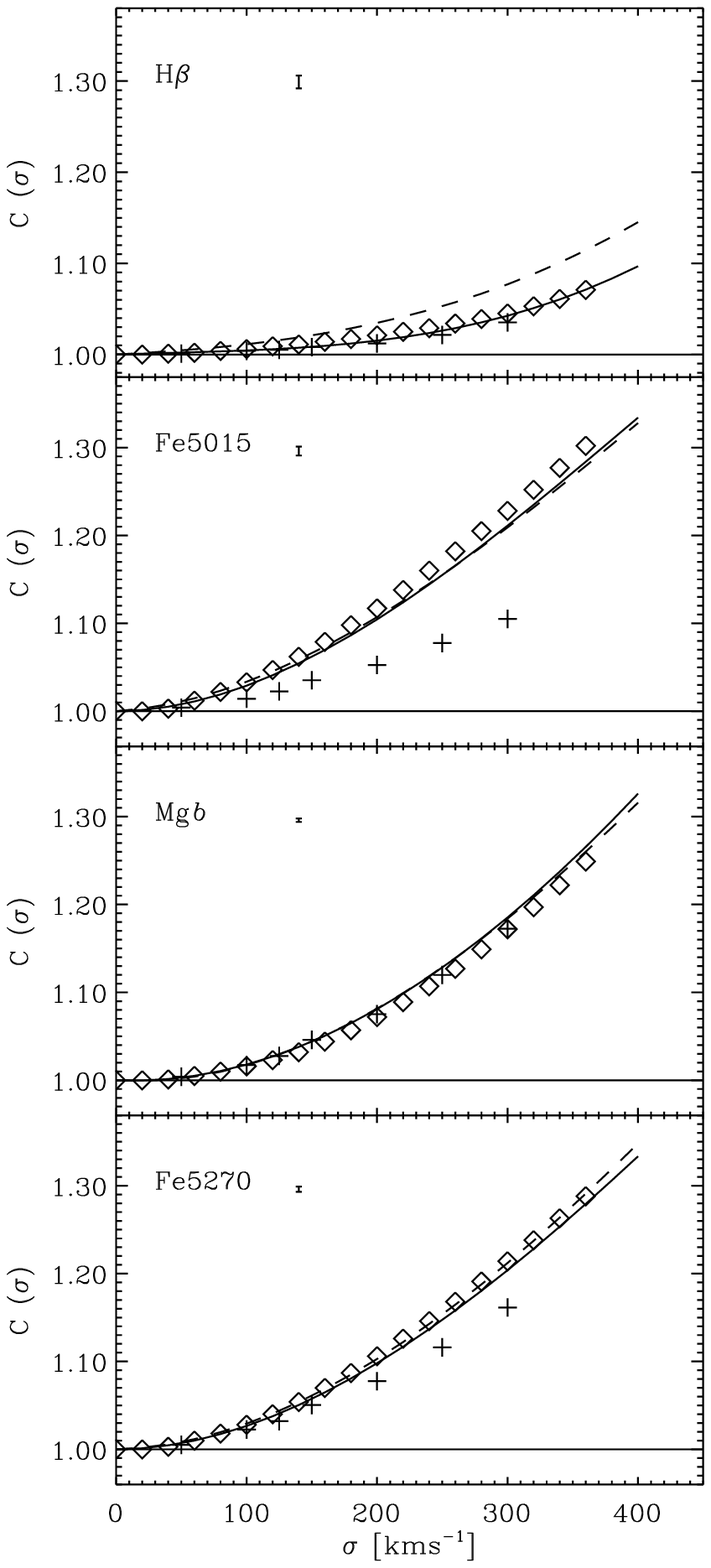}}
      \caption{Literature comparison of LOSVD correction curves for the 
        H$\beta$, Mg\,$b$\/ and Fe5270 indices. The dashed line shows
        the corrections derived by \citet{tra98}. The plus and open
        diamond signs stand for the \citet{long1998} and \citet{kun00}
        corrections, respectively. The parametrisation of this paper
        for purely Gaussian LOSVDs is given by the solid line. The
        error bar next to the index name represents the average
        $\pm1\sigma$ error in the correction factor due to differences
        in stellar populations (age $>$ 3\,Gyr) as investigated in this
        study.}
              \label{fig:lit_comp}%
\end{figure}

Before we start to discuss the results from this study, we present in
the following a comparison with the literature for purely Gaussian
LOSVDs. For the comparison we chose the calibrations by
\citet[][original Lick/IDS galaxy sample]{tra98}, \citet{long1998} and
\cite{kun00}. The literature calibrations were derived by using a
number of typically G or K giants as input spectra for the simulations.
For the indices in common we find good agreement in the general trends.
The agreement is of the order of $<$2\% for most of the indices between
this paper, \citet{tra98} and \citet{kun00}. Yet, for individual
indices such as H$\beta$ we find differences as large as 5\% (see
Figure~\ref{fig:lit_comp}). The agreement with the study of
\citet{long1998} is not as good and we find differences as large as
10\% at 300\,\kms\/ for the Fe5015 index (see
Figure~\ref{fig:lit_comp}). The disagreement for the H$\beta$ index
between \citet{tra98} and other studies can be largely explained by the
wide range of correction strengths derived from individual stars
\citep[for details see][]{kun00}. We believe that, by using model
spectra representative for early-type galaxies, this problem can be
overcome. The disagreement between \citet{long1998} and other studies
for the Fe5015 and Fe5270 indices is large and remains to be explained.
Overall, we conclude that the LOSVD corrections in the literature agree
qualitatively well, but significant quantitative differences remain.

\section{Results and discussion}
\label{sec:discussion}
\begin{figure*}
   \centering
   \includegraphics[width=14cm]{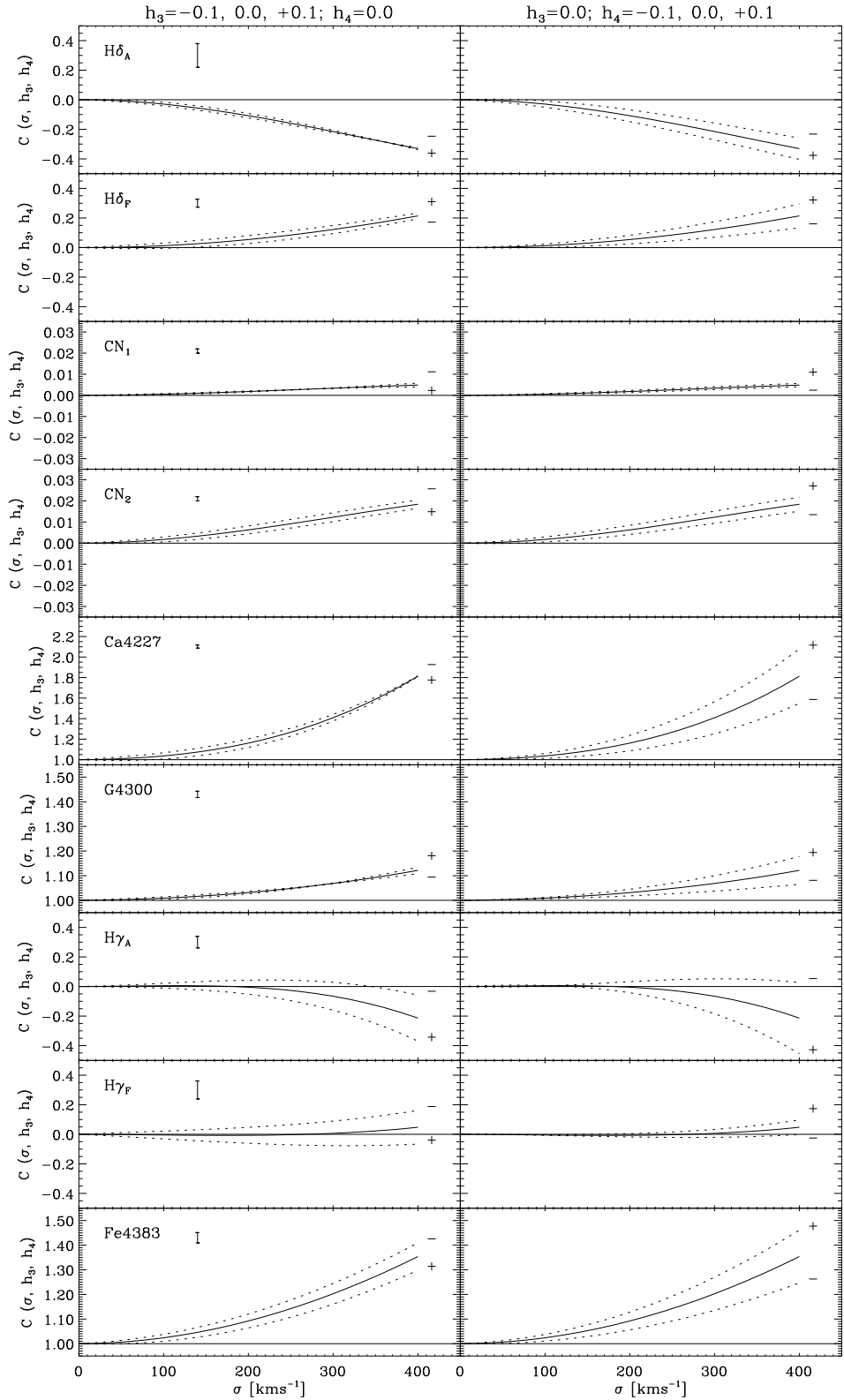}
      \caption{LOSVD corrections for old (age $\ge3$\,Gyr) stellar
        populations are shown for the following Lick/IDS indices:
        H$\delta_{\mathrm{A}}$, H$\delta_{\mathrm{F}}$, CN$_1$, CN$_2$,
        Ca4227, G4300, H$\gamma_{\mathrm{A}}$, H$\gamma_{\mathrm{F}}$,
        and Fe4383. The solid lines in all panels represent simulations
        for which $h_3=0.0$ and $h_4=0.0$.  The dotted lines show cases
        where either $h_3$ or $h_4$ is set to $-0.1$ and $+0.1$ as
        indicated by the plus and minus signs in the panels. The left
        panels show simulations for which $h_4=0.0$, whereas the right
        panels show simulations for $h_3=0.0$. The error bar next to
        the index name represents the average $\pm1\sigma$ error in the
        correction factor.}
              \label{fig:index_dispcor_blue}%
\end{figure*}

\begin{figure*}
   \centering
   \includegraphics[width=14cm]{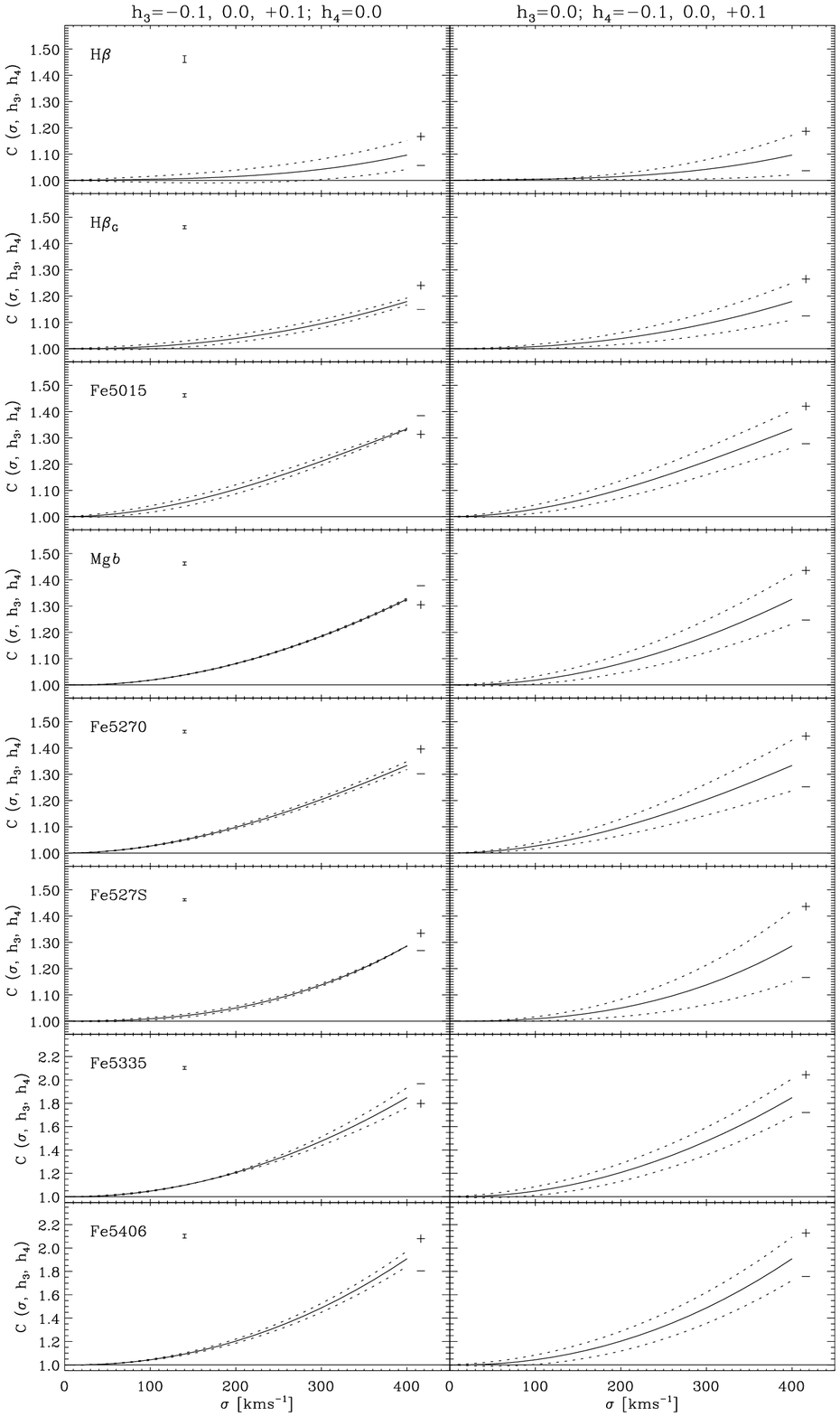}
      \caption{LOSVD corrections for old (age $\ge3$\,Gyr) stellar
        populations are shown for the following indices: H$\beta$,
        H$\beta_{\mathrm{G}}$, Fe5015, Mg\,$b$\/, Fe5270, Fe527S,
        Fe5335, and Fe5406. The solid lines in all panels represent
        simulations for which $h_3=0.0$ and $h_4=0.0$. The dotted lines
        show cases where either $h_3$ or $h_4$ is set to $-0.1$ and
        $+0.1$ as indicated by the plus and minus signs in the panels.
        The left panels show simulations for which $h_4=0.0$, whereas
        the right panels show simulations for $h_3=0.0$. The error bar
        next to the index name represents the average $\pm1\sigma$
        error in the correction factor.}
              \label{fig:index_dispcor_red}%
\end{figure*}

Most of the indices considered in this paper show significant changes
when the LOSVD is characterised by a non-zero $h_4$ term, while the
sensitivity to $h_3$ is negligible. Notable exceptions to this are the
H$\gamma_{\mathrm{A,F}}$, Fe4383, and H$\beta$ indices. These indices
are sensitive to both $h_3$ and $h_4$ terms.
Figures~\ref{fig:index_dispcor_blue} \& \ref{fig:index_dispcor_red}
schematically present for each index the dependency of the LOSVD
corrections on $\sigma$, $h_3$ and $h_4$.

Our simple correction model (Equations~\ref{equ:model_mult} \&
\ref{equ:model_add}) typically achieves a good fit to the set of galaxy
simulations with 1$\sigma$ errors of order 1.5\% or less for the
multiplicative corrections. For most of the investigations into the
stellar populations of early-type galaxies this accuracy is sufficient
since the error in the raw index measurements are typically larger.
For more accurate LOSVD corrections one needs to create an optimal
template of the average stellar population and kinematics of each
galaxy spectrum and use this individual template to derive the
corrections.

The simulated galaxy spectra, which we use to derive the LOSVD
corrections, are based on predicted SEDs for single-age,
single-metallicity stellar populations as provided by \citet{vaz99}.
Generally, these model spectra represent well the observed spectra of
early-type galaxies. However, the SEDs do not reflect, particularly at
solar metallicity and above, the super-solar $\alpha$-element to Fe
ratios typically found in elliptical and lenticular galaxies
\citep[e.g.,][]{WFG92,dav93,mar2003}. So, for individual indices
affected by non-solar abundance ratios, such as Mg\,$b$\/, the model
spectra often do not achieve a good fit. When model spectra with
varying abundance ratios become available, potential changes in the
LOSVD corrections will need to be investigated.

Most of the existing line-strengths studies in the Lick/IDS system did
not take into account the effects of non-Gaussian LOSVDs which raises
the question of how wrong their LOSVD corrections and interpretation of
the data were. An often used index combination to infer luminosity
weighted ages and metallicities of early-type galaxies is [MgFe] {\em
  vs}\/ H$\beta$~\footnote{${\mathrm {[MgFe]}} = {\sqrt{{\mathrm
        {Mg\,}}b \times ({\mathrm {Fe5270}} + {\mathrm {Fe5335}})/2}}$}
\citep[e.g.,][]{kun98,tra00a,meh03,TMB03}.  In order to assess the
effects on the derived stellar population parameters we use a simulated
galaxy spectrum representative of a central observation of a luminous
early-type galaxy with a large h$_4$ term ($\sigma = 250.0$\,\kms, h$_3
= 0.0$ and h$_4 = +0.10$; solar metallicity and 10\,Gyr of age). Using
the LICK/IDS system of line-strength indices and our newly proposed
LOSVD corrections we recover from the simulated galaxy spectrum an age
of 10.0\,Gyr and a metallicity of ${\mathrm {[Fe/H]}} = -0.01$. If we
would have used only the traditional corrections for the LOSVD
broadening, i.e.  measuring only a pure Gaussian velocity dispersion
($\sigma = 256.5$\,\kms) we would have obtained an age of 11.4\,Gyr and
a metallicity of ${\mathrm {[Fe/H]}} = -0.10$. The former exercise
assumes no errors in the determination of the line-strength indices and
kinematics.
  
For central galaxy line-strength measurements the observed LOSVDs
typically do not show such extreme h$_4$ terms and thus the traditional
LOSVD corrections, if well calibrated, are correct on the few percent
level. For line-strength gradients across galaxies one observes
significantly non-Gaussian LOSVDs \citep[e.g.,][]{em04} and therefore
the LOSVD corrections and thus the derived stellar population
parameters can indeed be significantly wrong.

\section{Conclusions}
\label{sec:conclusion}
In this paper we address the line-of-sight velocity distribution
(LOSVD) corrections for absorption line-strength indices of early-type
galaxies in the Lick/IDS system. We derive a simple parameterisation of
the corrections as a function of velocity dispersion $\sigma$ and the
first moments of the Gauss-Hermite series, $h_3$ and $h_4$, which
describe non-Gaussian LOSVDs. We provide calibrations for 15 Lick/IDS
indices and two additional indices.

For single stellar population ages greater than 3\,Gyr and
metallicities between [Fe/H] = $-0.7$ and $+0.2$ we find that LOSVD
corrections are almost independent of the stellar population. The
typical scatter is less than 1.5\% (rms). The LOSVD corrections for
higher order Balmer lines, which are important age indicators for the
study of early-type galaxies, show an increased sensitivity to stellar
population differences. However, the corrections themselves are small
compared to the typical range of line-strengths observed in early-type
galaxies. If more accurate corrections are needed, an optimal template
of the average stellar population and kinematics of each galaxy
spectrum can be created and used to derive precise corrections.

Investigating the sensitivity to non-Gaussian LOSVDs, parametrised by
the $h_3$ and $h_4$ terms of a Gauss-Hermite series, we find that the
H$\gamma_{\mathrm A,F}$, Fe4383, and H$\beta$ indices are sensitive to
both $h_3$ and $h_4$, while the remaining indices of this study are
sensitive to $h_4$ only. For variations of $h_4=\pm0.1$ at a constant
velocity dispersion of 250\,\kms\/ we typically find changes of
$\pm5\%$ in the LOSVD correction. These changes translate into
approximately 15\% and 20\% errors in the ages and metallicities of old
stellar populations, respectively, as estimated from index-index
diagrams.

\begin{acknowledgements}
  We thank R. McDermid, C. Harrison, E. Emsellem, M. Cappellari and the
  referee S. Trager for useful discussions and advice.
\end{acknowledgements}

\bibliographystyle{aa}
\bibliography{references}

\begin{thebibliography}{21}
\expandafter\ifx\csname natexlab\endcsname\relax\def\natexlab#1{#1}\fi

\bibitem[{{Bender} {et~al.}(1994){Bender}, {Saglia}, \& {Gerhard}}]{ben94}
{Bender}, R., {Saglia}, R.~P., \& {Gerhard}, O.~E. 1994, \mnras, 269, 785

\bibitem[{{Cappellari} \& {Emsellem}(2004)}]{cap04}
{Cappellari}, M. \& {Emsellem}, E. 2004, \pasp, 116, 138

\bibitem[{{Davies} {et~al.}(1993){Davies}, {Sadler}, \& {Peletier}}]{dav93}
{Davies}, R.~L., {Sadler}, E.~M., \& {Peletier}, R.~F. 1993, \mnras, 262, 650

\bibitem[{{Emsellem} {et~al.}(2004){Emsellem}, {Cappellari}, {Peletier},
  {McDermid}, {Bacon}, {Bureau}, {Copin}, {Davies}, {Krajnovi\'{c}},
  {Kuntschner}, {Miller}, \& {de Zeeuw}}]{em04}
{Emsellem}, E., {Cappellari}, M., {Peletier}, R.~F., {et~al.} 2004, \mnras, in
  press

\bibitem[{{Faber} {et~al.}(1985){Faber}, {Friel}, {Burstein}, \&
  {Gaskell}}]{fab85}
{Faber}, S.~M., {Friel}, E.~D., {Burstein}, D., \& {Gaskell}, C.~M. 1985,
  \apjs, 57, 711

\bibitem[{{Gerhard}(1993)}]{ger93}
{Gerhard}, O.~E. 1993, \mnras, 265, 213

\bibitem[{{J{\o}rgensen}(1997)}]{jor97}
{J{\o}rgensen}, I. 1997, \mnras, 288, 161

\bibitem[{{Kuntschner}(2000)}]{kun00}
{Kuntschner}, H. 2000, \mnras, 315, 184

\bibitem[{{Kuntschner} \& {Davies}(1998)}]{kun98}
{Kuntschner}, H. \& {Davies}, R.~L. 1998, \mnras, 295, L29+

\bibitem[{{Longhetti} {et~al.}(1998){Longhetti}, {Rampazzo}, {Bressan}, \&
  {Chiosi}}]{long1998}
{Longhetti}, M., {Rampazzo}, R., {Bressan}, A., \& {Chiosi}, C. 1998, \aaps,
  130, 251

\bibitem[{{Maraston} {et~al.}(2003){Maraston}, {Greggio}, {Renzini},
  {Ortolani}, {Saglia}, {Puzia}, \& {Kissler-Patig}}]{mar2003}
{Maraston}, C., {Greggio}, L., {Renzini}, A., {et~al.} 2003, \aap, 400, 823

\bibitem[{{Mehlert} {et~al.}(2003){Mehlert}, {Thomas}, {Saglia}, {Bender}, \&
  {Wegner}}]{meh03}
{Mehlert}, D., {Thomas}, D., {Saglia}, R.~P., {Bender}, R., \& {Wegner}, G.
  2003, \aap, 407, 423

\bibitem[{{Scorza} \& {Bender}(1995)}]{sco95}
{Scorza}, C. \& {Bender}, R. 1995, \aap, 293, 20

\bibitem[{{Thomas} {et~al.}(2003){Thomas}, {Maraston}, \& {Bender}}]{TMB03}
{Thomas}, D., {Maraston}, C., \& {Bender}, R. 2003, \mnras, 339, 897

\bibitem[{{Trager} {et~al.}(2000){Trager}, {Faber}, {Worthey}, \& {Gonz{\'
  a}lez}}]{tra00a}
{Trager}, S.~C., {Faber}, S.~M., {Worthey}, G., \& {Gonz{\' a}lez}, J.~J. 2000,
  \aj, 119, 1645

\bibitem[{{Trager} {et~al.}(1998){Trager}, {Worthey}, {Faber}, {Burstein}, \&
  {Gonzalez}}]{tra98}
{Trager}, S.~C., {Worthey}, G., {Faber}, S.~M., {Burstein}, D., \& {Gonzalez},
  J.~J. 1998, \apjs, 116, 1

\bibitem[{{van der Marel} \& {Franx}(1993)}]{mf93}
{van der Marel}, R.~P. \& {Franx}, M. 1993, \apj, 407, 525

\bibitem[{{Vazdekis}(1999)}]{vaz99}
{Vazdekis}, A. 1999, \apj, 513, 224

\bibitem[{{Worthey} {et~al.}(1992){Worthey}, {Faber}, \& {Gonzalez}}]{WFG92}
{Worthey}, G., {Faber}, S.~M., \& {Gonzalez}, J.~J. 1992, \apj, 398, 69

\bibitem[{{Worthey} {et~al.}(1994){Worthey}, {Faber}, {Gonzalez}, \&
  {Burstein}}]{wor94b}
{Worthey}, G., {Faber}, S.~M., {Gonzalez}, J.~J., \& {Burstein}, D. 1994, ApJS,
  94, 687

\bibitem[{{Worthey} \& {Ottaviani}(1997)}]{worott97}
{Worthey}, G. \& {Ottaviani}, D.~L. 1997, ApJS, 111, 377

\end{thebibliography}

\end{document}